\documentclass[10pt,letterpaper]{article}
\usepackage[top=0.85in,left=2.75in,footskip=0.75in]{geometry}

\usepackage{amsmath,amssymb}

\usepackage{changepage}

\usepackage[utf8x]{inputenc}

\usepackage{textcomp,marvosym}

\usepackage{cite}

\usepackage{nameref}

\usepackage[right]{lineno}

\usepackage{microtype}
\DisableLigatures[f]{encoding = *, family = * }

\usepackage[table]{xcolor}

\usepackage{array}

\newcolumntype{+}{!{\vrule width 2pt}}

\newlength\savedwidth

\newcommand\thickhline{\noalign{\global\savedwidth\arrayrulewidth\global\arrayrulewidth 2pt}%
\hline
\noalign{\global\arrayrulewidth\savedwidth}}

\raggedright
\usepackage[aboveskip=1pt,labelfont=bf,labelsep=period,justification=raggedright,singlelinecheck=off]{caption}

\makeatletter
\renewcommand{\@biblabel}[1]{\quad#1.}
\makeatother

\usepackage{lastpage,fancyhdr,graphicx}
\usepackage{epstopdf}
\fancyheadoffset[L]{2.25in}
\fancyfootoffset[L]{2.25in}
\usepackage{hhline}
\usepackage{multirow}

\usepackage{tikz}
\usetikzlibrary{shapes}
\newcommand{\ldata}{\raisebox{2pt}{\tikz{\draw[-,black!50!green,dashed,line width=0.75pt] (0,0) --
(4mm,0);}}}
\newcommand{\leqn}{\raisebox{2pt}{\tikz{\draw[-,orange,dotted,line width=0.75pt] (0,0) --
(4mm,0);}}} 
\newcommand{\lhybrid}{\raisebox{2pt}{\tikz{\draw[-,blue,dashdotted,line width=0.75pt] (0,0) --
(4mm,0);}}}
\newcommand{\ltruth}{\raisebox{2pt}{\tikz{\draw[-,black,solid,line width=0.75pt] (0,0) --
(4mm,0);}}}
\newcommand{\pdata}{\tikz{\node[draw=black!50!green,fill=black!50!green,rectangle,minimum
width=0.15cm,minimum height=0.15cm,inner sep=0pt] at (0,0) {};}}
\newcommand{\phybrid}{\tikz{\node[draw=blue,fill=blue,circle,minimum
width=0.15cm,minimum height=0.15cm,inner sep=0pt] at (0,0) {};}}
\newcommand{\peqn}{\tikz{\node[draw=orange,fill=orange,isosceles
triangle,isosceles triangle stretches,shape border rotate=90,minimum
width=0.15cm,minimum height=0.15cm,inner sep=0pt] at (0,0) {};}}

\begin{document}
\vspace*{0.2in}

\begin{flushleft}
{\Large
\textbf\newline{Data-assisted reduced-order modeling of extreme events in complex dynamical systems} 
}
\newline
\\
Zhong Y. Wan\textsuperscript{1},
Pantelis R. Vlachas\textsuperscript{2},
Petros Koumoutsakos\textsuperscript{2},
Themistoklis P. Sapsis\textsuperscript{1*}
\\
\bigskip
\textbf{1} Department of Mechanical Engineering, Massachusetts Institute of Technology, Cambridge, MA, USA
\\
\textbf{2} Chair of Computational Science, ETH Zurich, Zurich, Switzerland
\\
\bigskip

%
%

* Corresponding author. Email: sapsis@mit.edu (TPS)

\end{flushleft}
\section*{Abstract}
\noindent The prediction of extreme events, from avalanches and droughts to tsunamis and epidemics, depends on the formulation and analysis of relevant, complex dynamical systems. Such dynamical systems are characterized by high intrinsic dimensionality with  extreme events having the form of rare transitions that are several standard deviations away from the mean. Such systems are not amenable to classical order-reduction methods through projection of the governing equations due to the large intrinsic dimensionality of the underlying attractor as well as  the complexity of the transient events. Alternatively, data-driven techniques aim to quantify the dynamics of specific, critical modes by utilizing data-streams and  by expanding the dimensionality of the reduced-order model using delayed coordinates. In turn, these methods have major limitations in regions of the phase space with sparse data, which is the case for extreme events.
	
In this work, we develop a novel hybrid framework that complements an imperfect reduced order model, with data-streams that are integrated though a recurrent neural network (RNN) architecture. The reduced order model has the form of projected equations into a low-dimensional subspace that still contains important dynamical information about the system and it is expanded by a long short-term memory (LSTM) regularization. The LSTM-RNN is trained by analyzing the mismatch between the imperfect model and the data-streams, projected to the reduced-order space. The  data-driven model assists the imperfect model in regions where data is available, while for locations where data is sparse  the imperfect model still provides a baseline for the prediction of the system state. We assess the developed framework on two challenging prototype systems exhibiting extreme events. We  show that the blended approach has improved performance compared with methods that use either data streams or the imperfect model alone. Notably the  improvement is more significant in regions associated with extreme events, where data is sparse.


\section*{Introduction}

Extreme events are omnipresent in important problems in science and technology such as turbulent and reactive flows
\cite{smooke86, pope97}, Kolmogorov \cite{chandler13} and unstable plane Couette flow
\cite{hamilton95}), geophysical systems (e.g. climate dynamics \cite{majda11,
majda2000}, cloud formations in tropical atmospheric convection \cite{grab99, grab01}), nonlinear
optics \cite{akhm13, arec11} or water waves \cite{onorato13, muller, fedele2008}),
and mechanical systems (e.g. mechanical metamaterials \cite{kim15, li14}).

The complete description of these system through the governing equations is often
challenging either because it is very hard/expensive to solve these equations with an appropriate 
resolution or due to the magnitude of the model errors. The very large dimensionality
of their attractor in combination with the occurrence of important transient, but rare events, makes
the application of classical order-reduction methods a challenging task. Indeed, classical Galerkin
projection methods  encounter problems  as the truncated
degrees-of-freedom are often essential for the effective description of the system due to high
underlying intrinsic dimensionality. On the other hand, purely data-driven, non-parametric methods
such as delay embeddings \cite{farmer87, crutchfield87, sugihara90, rowlands92, Berry2015,
Berry2016}, equation-free methods \cite{kevrekidis03, kevrekidis05}, Gaussian process regression
based methods \cite{Zhong16}, or recurrent neural networks based approaches \cite{Vlachas18} may
not perform well when it comes to rare events, since the training data-sets typically contain only
a small number of the rare transient responses. The same limitations hold for data-driven,
parametric methods \cite{Bongard2007, Schmidt09, brunton16, raissi17}, where the assumed analytical
representations have parameters that are optimized so that the resulted model best fits the data.
Although these methods perform well when the system operates within the main `core' of the
attractor, this may not be the case when rare and/or extreme events occur.

We propose a hybrid method for the formulation of a reduced-order model that combines  an imperfect physical model with available data streams. The proposed framework is
important for the non-parametric description, prediction and control of complex systems whose
response is characterized by both i) high-dimensional attractors with broad energy spectrum
distributed across multiple scales, and ii) strongly transient non-linear dynamics such as extreme
events. 

We focus on data-driven recurrent neural networks (RNN) with a long-short term memory (LSTM)
\cite{LSTM} that represents some of the truncated degrees-of-freedom. The key concept of our work is
the observation that while the imperfect model alone has limited  descriptive and prediction skills (either because it has been obtained by a radical reduction or it is a coarse-grid solution of the original equations), it still contains important information
especially for the instabilities of the system, assuming that the relevant modes are included in
the truncation. However, these instabilities need to be combined with an accurate description of
the nonlinear dynamics within the attractor and this part is captured in the present framework
by the recurrent neural network. Note, that embedding theorems \cite{Whitney36, Takens81} make the
additional memory of the RNN to represent dimensions of the system that have been truncated, a
property that provides an additional advantage in the context of reduced-order modeling
\cite{Berry2015, Vlachas18}.

We note that such blended model-data approaches have been proposed previously in other contexts. 
In \cite{Ott18, Ott18-2}, a hybrid forecasting scheme based on reservoir computing in conjunction with 
knowledge-based models are successfully applied to example high-dimensional spatiotemporal chaotic systems.
In \cite{sapsis_majda_tur, sapsis_majda_mqg, sapsis_majda_mqgdo} the linearized
dynamics were projected to low-dimensional subspaces and were combined with additive noise and
damping that were rigorously selected to represent the effects nonlinear energy fluxes from the
truncated modes. The developed scheme resulted in reduced-order stochastic models that efficiently
represented the second order statistics in the presence of arbitrary external excitation. In
\cite{Milano2002} a deep neural network architecture was developed to reconstruct the near-wall
flow field in a turbulent channel flow using suitable wall only information. These nonlinear
near-wall models can be integrated with flow solvers for the parsimonious modeling and control of
turbulent  flows\cite{Bright:2013,Ling2016a,Guniat2016,Wang:2017}. In \cite{Weymouth2014} a
framework was introduced wherein solutions from intermediate models, which capture some physical
aspects of the problem, were incorporated as solution representations into machine learning tools
to improve the predictions of the latter, minimizing the reliance on costly experimental
measurements or high-resolution, high-fidelity numerical solutions. \cite{Sanner1995} design a
stable adaptive control strategy using neural networks for physical systems for which the state
dependence of the dynamics is reasonably well understood, but the exact functional form of this
dependence, or part thereof, is not, such as underwater robotic vehicles and high performance
aircraft. In \cite{Rico-Martin:1994,Lagaris1998,raissi17,raissi18} neural nets are developed to
simultaneously learn the solution of the model equations  using data. In these works that only a
small number of scalar parameters is utilized to represent unknown dynamics, while the emphasis is
given primarily on the learning of the solution, which is represented through a deep neural
network. In other words, it is assumed that a family of models that `lives' in a low-dimensional
parameter space can capture the correct response. Such a representation is not always available
though. Here our goal is to apply such a philosophy on the prediction of complex systems
characterized by high dimensionality and strongly transient dynamics. We demonstrate the developed
strategy in prototype systems exhibiting extreme events and show that the hybrid strategy has
important advantages compared with either purely data-driven methods or those relying on
reduced-order models alone.

\section*{Materials and methods}

We consider a nonlinear dynamical system with state variable $\mathbf{u} \in \mathbb{R}^d$ and
dynamics given by
\begin{equation}
	\frac{d\mathbf{u}}{dt} = \mathbf{F}(\mathbf{u}) = \mathbf{L}\mathbf{u} + \mathbf{h}(\mathbf{u}),
	\label{eq:dynamical_system}
\end{equation} where $\mathbf{F}:\mathbb{R}^d\rightarrow\mathbb{R}^d$ is a deterministic,
time-independent operator with linear and nonlinear parts $\mathbf{L}$ and $\mathbf{h}$
respectively. We are specifically interested in systems whose dynamics results in a non-trivial,
globally attracting manifold $S \subset \mathbb{R}^d$ to which trajectories quickly decay. The
intrinsic dimension of $S$ is presumably much less than $d$.

In traditional Galerkin-based reduced-order model \cite{Matthies03} one typically uses an ansatz of
the form
\begin{equation}
	\mathbf{u} = \mathbf{Y}\boldsymbol{\xi} + \mathbf{Z}\boldsymbol{\eta} + \mathbf{b},
	\label{eq:ansatz}
\end{equation} where the columns of matrix $\mathbf{Y} = [\mathbf{y}_1, ..., \mathbf{y}_m]$
form an orthonormal basis of $Y$, an $m$-dimensional subspace of $\mathbb{R}^d$, and the
columns of $\mathbf{Z} = [\mathbf{z}_{1},...,\mathbf{z}_{d-m}]$ make up an orthonormal basis for the
orthogonal complement $Z = \mathbb{R}^d \setminus Y$; $\boldsymbol{\xi}$ and $\boldsymbol{\eta}$ are
the projection coordinates associated with $\mathbf{Y}$ and $\mathbf{Z}$; $\mathbf{b}$ is an offset
vector typically made equal to the attractor mean state. This linear expansion allows reduction to
take place through special choices of subspace $Y$ and $Z$, as well as their corresponding basis.
For example, the well-known proper orthogonal decomposition (POD) derives the subspace empirically
to be such that the manifold $S$ preserves its variance as much as possible when projected to $Y$ (or
equivalently, minimizing the variance when projected to $Z$), given a fixed dimension constraint $m$.

We show that such a condition enables reduction, by substituting Eq~(\ref{eq:ansatz}) into
Eq~(\ref{eq:dynamical_system}) and projecting onto $Y$ and $Z$ respectively to obtain two coupled systems
of differential equations:
\begin{equation} 
	\begin{gathered}
		\frac{d\boldsymbol{\xi}}{dt} = \mathbf{Y}^T\mathbf{L}\mathbf{Y}\boldsymbol{\xi} +
			\mathbf{Y}^T\mathbf{L}\mathbf{Z}\boldsymbol{\eta} + 
			\mathbf{Y}^T\mathbf{h}(\mathbf{Y}\boldsymbol{\xi} + \mathbf{Z}\boldsymbol{\eta} + \mathbf{b}) +
			\mathbf{Y}^T\mathbf{L}\mathbf{b} \\
		\frac{d\boldsymbol{\eta}}{dt} = \mathbf{Z}^T\mathbf{L}\mathbf{Y}\boldsymbol{\xi} +
			\mathbf{Z}^T\mathbf{L}\mathbf{Z}\boldsymbol{\eta} +
			\mathbf{Z}^T\mathbf{h}(\mathbf{Y}\boldsymbol{\xi} + \mathbf{Z}\boldsymbol{\eta} + \mathbf{b}) +
			\mathbf{Z}^T\mathbf{L}\mathbf{b}.
	\end{gathered} 
	\label{eq:coupled} 
\end{equation} If on average $|\boldsymbol{\eta}| \ll |\boldsymbol{\xi}|$, we may make the
approximation that $\boldsymbol{\eta} = 0$, leading to a $m$-dimensional system (ideally $m \ll d$)
\begin{equation}
	\frac{d\boldsymbol{\xi}}{dt} = \mathbf{Y}^T\mathbf{L}\mathbf{Y}\boldsymbol{\xi} +
		\mathbf{Y}^T\mathbf{h}(\mathbf{Y}\boldsymbol{\xi} + \mathbf{b}) + \mathbf{Y}^T\mathbf{L}\mathbf{b}
		= \mathbf{F}_{\boldsymbol{\xi}}(\boldsymbol{\xi}),
	\label{eq:flat_Galerkin}
\end{equation} which can be integrated in time. This is known as the \textit{flat Galerkin method}.
The solution to Eq~(\ref{eq:dynamical_system}) is approximated by $\mathbf{u}\approx\mathbf{Y}\boldsymbol{\xi} + \mathbf{b}$.

Using (\ref{eq:flat_Galerkin}) as an approximation to Eq~(\ref{eq:dynamical_system}) is known to suffer
from a number of problems. First, the dimension $m$ of the reduction subspace $Y$ may be too large
for $|\boldsymbol{\eta}|\approx 0$ to hold true. Second, the subspace $Z$ is derived merely based on
statistical properties of the manifold without addressing the dynamics. This implies
that even if $\boldsymbol{\eta}$ has small magnitude on average it may play a big role in
the dynamics of the high-energy space (e.g. acting as buffers for energy transfer between
modes \cite{sapsis_majda_qgdo}). Neglecting such dimensions in the description of the system may alter its dynamical behaviors and compromise the ability of the model to generate reliable forecasts.

\begin{figure}[!b]
\centering
\includegraphics[width=\textwidth]{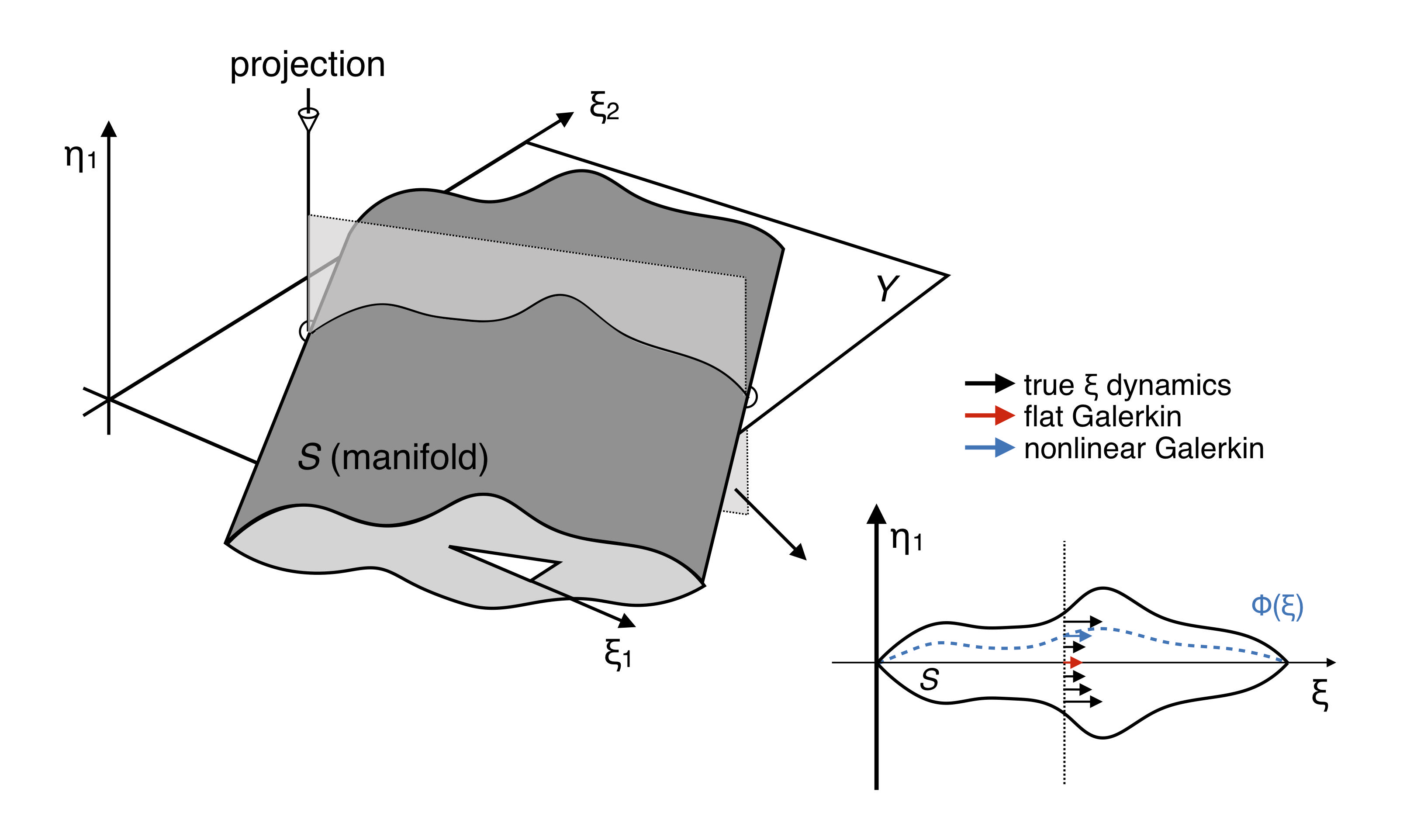}
\caption{{\bf Geometric illustration of flat and nonlinear Galerkin projected dynamics in $\mathbb{R}^3$.} 3D manifold $S$ living in $(\xi_1,\xi_2, \eta_1)$ is projected to 2D plane parametrized by $(\xi_1,\xi_2)$. Parametrization is assumed to be imperfect, i.e. out-of-plane coordinate $\eta_1$ cannot be uniquely determined from $(\xi_1,\xi_2)$. Flat Galerkin method always uses the dynamics corresponding to $\eta_1 = 0$. Nonlinear Galerkin method uses the dynamics corresponding to $\eta_1 = \Phi(\xi_1, \xi_2)$ where $\Phi$ is determined by some prescribed criterion (e.g. minimization of $L2$ error). }
\label{fig1}
\end{figure}

An existing method that attempts to address the truncation effect of the $\boldsymbol{\eta}$
terms is the \textit{nonlinear Galerkin projection} \cite{Matthies03, Foias88}, which expresses
$\boldsymbol{\eta}$ as a function of $\boldsymbol{\xi}$:
\begin{equation}
	\boldsymbol{\eta} = \boldsymbol{\Phi}(\boldsymbol{\xi}),
	\label{eq:nlGalerkin_assumption}
\end{equation} yielding a reduced system
\begin{equation}
	\frac{d\boldsymbol{\xi}}{dt} = \mathbf{Y}^T\mathbf{L}\mathbf{Y}\boldsymbol{\xi} +
	\mathbf{Y}^T\mathbf{LZ}\boldsymbol{\Phi}(\boldsymbol{\xi}) +
	\mathbf{Y}^T\mathbf{h}(\mathbf{Y}\boldsymbol{\xi} + \mathbf{Z}\boldsymbol{\Phi}(\boldsymbol{\xi})
	+ \mathbf{b}) + \mathbf{Y}^T\mathbf{L}\mathbf{b}.
	\label{eq:nl_Galerkin}
\end{equation} The problems boils down to finding $\boldsymbol{\Phi}$, often empirically.
Unfortunately, $\boldsymbol{\Phi}$ is well-defined only when the inertial manifold $S$ is fully
parametrized by dimensions of $Y$ (see Fig~\ref{fig1}), which is a difficult condition
to achieve for most systems under a reasonable $m$. Even if the condition is met, how to
systematically find $\boldsymbol{\Phi}$ remains a big challenge.


\subsection*{Data-assisted reduced-order modeling}

In this section we introduce a new framework for improving the reduced-space model
that assists, with data streams, the nonlinear Galerkin method. Our main idea relies on building an additional
data-driven model from data series observed in the reduction space to assist the equation-based
model equations (\ref{eq:flat_Galerkin}).

We note  that the \textit{exact} dynamics of $\boldsymbol{\xi}$ can be written as
\begin{equation}
	\frac{d\boldsymbol{\xi}}{dt} = \mathbf{F}_{\boldsymbol{\xi}}(\boldsymbol{\xi}) +
		\mathbf{G}(\boldsymbol{\xi}, \boldsymbol{\eta}),
	\label{eq:exact_red_dynamics}
\end{equation} where $\mathbf{F}_{\boldsymbol{\xi}}$ is defined in Eq~(\ref{eq:flat_Galerkin}) and
$\mathbf{G}:\mathbb{R}^m\times\mathbb{R}^{d-m}\rightarrow\mathbb{R}^m$ encompasses the
coupling between $\boldsymbol{\xi}$ and $\boldsymbol{\eta}$. We will refer to
$\boldsymbol{\psi} = \mathbf{G}(\boldsymbol{\xi}, \boldsymbol{\eta})$ as the
\textit{complementary dynamics} since it can be thought of as a correction that complements the
flat Galerkin dynamics $\mathbf{F}_{\boldsymbol{\xi}}$.

The key step of our framework is to establish a data-driven model $\hat{\mathbf{G}}$ to
approximate $\mathbf{G}$:
\begin{equation}
	\boldsymbol{\psi}(t) = \mathbf{G}\bigg(\boldsymbol{\xi}(t),
		\boldsymbol{\eta}(t)\bigg)\approx\hat{\mathbf{G}}\bigg(\boldsymbol{\xi}(t),
		\boldsymbol{\xi}(t-\tau), \boldsymbol{\xi}(t-2\tau),...\bigg)
	\label{eq:approx_comp_dynamics}
\end{equation} where $\boldsymbol{\xi}(t), \boldsymbol{\xi}(t-\tau),
...$ are uniformly time-lagged states in $\boldsymbol{\xi}$ up to a reference initial condition.
The use of delayed $\boldsymbol{\xi}$ states makes up for the fact that $Y$ may not be a
perfect parametrization subspace for $S$. The missing state information not directly
accessible from within $Y$ is instead inferred from these delayed $\boldsymbol{\xi}$ states and
then used to compute $\boldsymbol{\psi}$. This model form is motivated by the embedding
theorems developed by Whitney \cite{Whitney36} and Takens \cite{Takens81}, who showed
that the attractor of a deterministic, chaotic dynamical system can be fully embedded using delayed
coordinates.

We use the long short-term memory (LSTM) \cite{LSTM}, a regularization of recurrent neural network (RNN),
as the fundamental building block for constructing $\hat{G}$. The LSTM has been recently deployed
successfully for the formulation of fully data-driven models for the prediction of complex
dynamical systems \cite{Vlachas18}. Here we employ the same strategy to model the complementary
dynamics while we preserve the structure of the projected equations. LSTM takes advantage of the
sequential nature of the time-delayed reduced space coordinates by processing the input in
chronological order and keeping memory of the useful state information that complements
$\boldsymbol{\xi}$ at each time step. An overview of the RNN model and the LSTM is given in
\nameref{S1_appendix}.

Building from LSTM units, we use two different architectures to learn the complementary
dynamics from data. The first architecture reads a sequence of $\boldsymbol{\xi}$ states, i.e.
states projected to the $d-$dimensional subspace and outputs the corresponding sequence of
complementary dynamics. The second architecture reads an input sequence and integrate the output
dynamics to predict future. The details of both architectures are described below.

\subsubsection*{Data series}

Both architectures are trained and tested on the same data set consisting of $N$ data series, where
$N$ is assumed to be large enough such that the low-order statistics of $S$ are accurately
represented. Each data series is a sequence of observed values in reduced space $Y$, with strictly
increasing and evenly spaced observation times. Without loss of generality, we assume that all data series 
have the same length. Moreover, the observation time spacing $\tau$ is assumed to be small so that the 
true dynamics at each step of the series can be accurately estimated with finite difference.  We remark 
that for single-step prediction (architecture I below) increasing $\tau$ (while keeping the number of steps 
constant) is beneficial for training as it reduces the correlation between successive inputs. However, for 
multi-step prediction (architecture II), large $\tau$ incurs integration errors which quickly outweigh the benefit of having decorrelated inputs. Hence, we require small $\tau$ in data.

\subsubsection*{Architecture I}

\begin{figure}[!h]
\centering
\includegraphics[width=\textwidth]{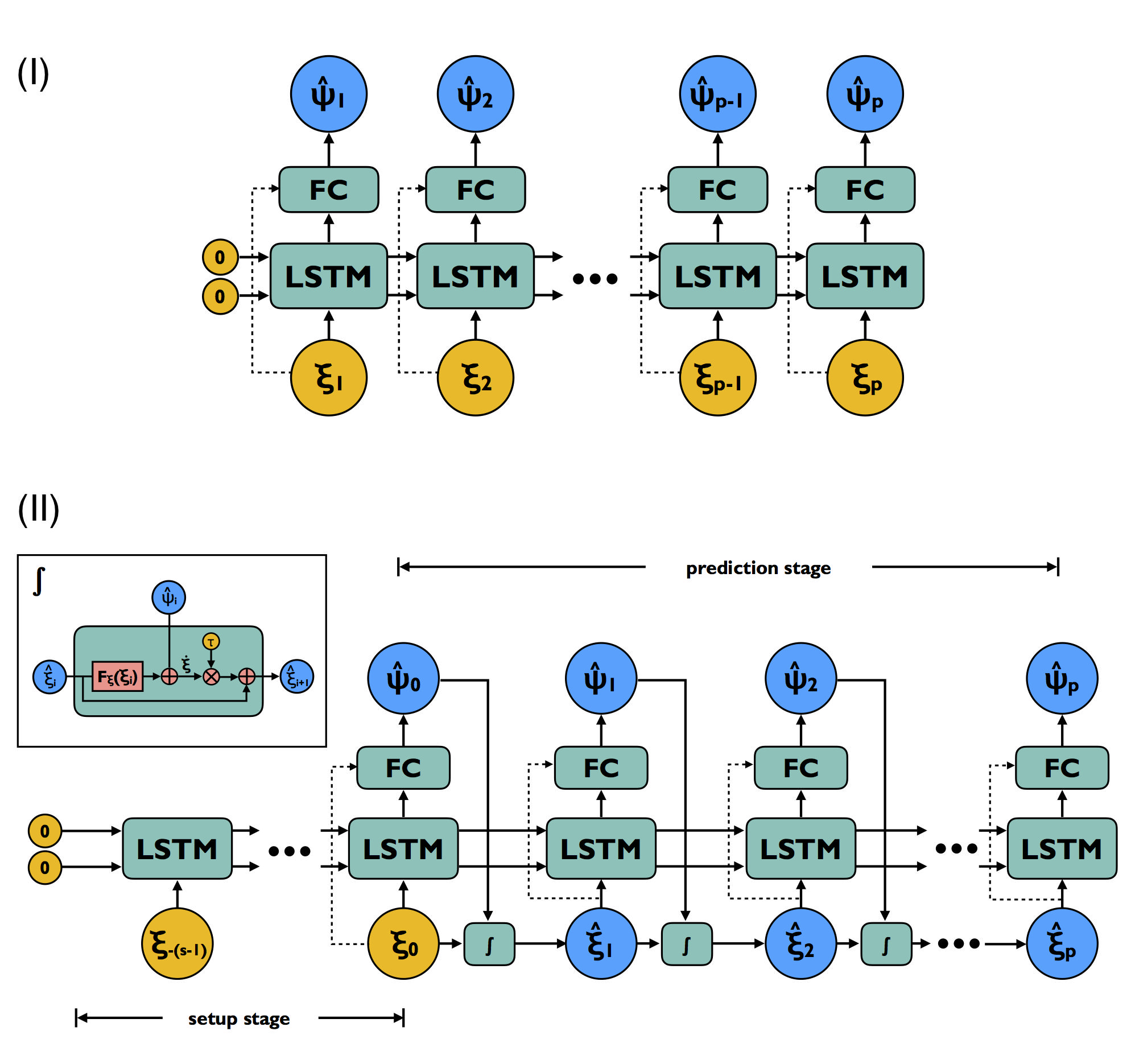}
\caption{{\bf Computational graph for model architecture I and II.} Yellow nodes are input provided to the network corresponding to sequence of states and blue nodes are prediction targets corresponding to the complementary dynamics (plus states for architecture II). Blocks labeled `FC' are fully-connected layers with ReLU activations. Dashed arrows represent optional connections depending on the capacity of LSTM relative to the dimension of $\boldsymbol{\xi}$. Both architectures share the same set of trainable weights. For architecture I, predictions are made as input is read; input is always accurate regardless of any prediction errors made in previous steps. This architecture is used \textit{only for training}. Architecture II makes prediction in a sequence-to-sequence (setup sequence to prediction sequence) fashion. Errors made early do impact all predictions that follow. This architecture is used for \textit{fine-tuning weights} and \textit{multi-step-ahead prediction}.}
\label{fig2}
\end{figure}

We denote an input sequence of length-$p$ as $\{\boldsymbol{\xi}_1, ..., \boldsymbol{\xi}_{p}\}$ and
the corresponding finite-difference interpolated dynamics as $\{\dot{\boldsymbol{\xi}}_1, ...,
\dot{\boldsymbol{\xi}}_{p}\}$. A forward pass in the first architecture works as follows
(illustrated in Fig~\ref{fig2}I). At time step $i$, input $\boldsymbol{\xi}_i$ is fed
into a LSTM cell with $n_{\text{LSTM}}$ hidden states, which computes its output $\mathbf{h}_i$
based on the received input and its previous memory states (initialized to zero). The LSTM output
is then passed through an intermediary fully-connected (FC) layer with $n_{\text{FC}}$ hidden
states and rectified linear unit (ReLU) activations to the output layer at desired dimension $m$.
Here $\mathbf{h}_i$ is expected to contain state information of the \textit{unobserved}
$\boldsymbol{\eta}$ at time step $i$, reconstructed effectively as a function of all previous
\textit{observed} states $\{\boldsymbol{\xi}_1, \ldots, \boldsymbol{\xi}_{i-1}\}$. The model output
is a predicted sequence of complementary dynamics $\{\hat{\boldsymbol{\psi}}_1, \ldots,
\hat{\boldsymbol{\psi}}_p\}$.


Optionally, $\mathbf{h}_i$ can be concatenated with LSTM input $\boldsymbol{\xi}_i$ to make up the
input to the FC layer. The concatenation is necessary when $n_{\text{LSTM}}$ is small relative to
$m$. Under such conditions, the LSTM hidden states $\mathbf{h}_i$ are more likely trained to
represent $\boldsymbol{\eta}_i$ alone, as opposed to $(\boldsymbol{\xi}_i, \boldsymbol{\eta}_i)$
combined. $\boldsymbol{\xi}_i$ thus needs to be seen by the FC layer in order to have all elements
necessary in order to estimate $\boldsymbol{\psi}$. If the LSTM cell has sufficient room to
integrate incoming input with memory (i.e. number of hidden units larger than the intrinsic
dimensionality of the attractor), the concatenation may be safely ignored.

In the case of models with lower complexity for faster learning a small $n_{\text{LSTM}}$ is preferred. However,
finding the minimum working $n_{\text{LSTM}}$, which is expected to approach the intrinsic attractor
dimension, is a non-trivial problem. Therefore, it is sometimes desirable to
conservatively choose $n_{\text{LSTM}}$. In this case the LSTM unit is likely to have sufficient
cell capacity to integrate incoming input with memory, rendering the input concatenation step unnecessary.

The model is trained by minimizing a loss function with respect to the weights of the LSTM cell and
FC layer. The loss function is defined as a weighted sum of the mean squared error (MSE) of the
complementary dynamics:

\begin{equation}
	L = \sum\limits_{i=1}^p w_i||\hat{\boldsymbol{\psi}}_i - \boldsymbol{\psi}_i||^2
	\label{eq:archI_loss}
\end{equation} where $\boldsymbol{\psi}_i = \dot{\boldsymbol{\xi}}_i -
\mathbf{F}_{\boldsymbol{\xi}}(\boldsymbol{\xi})$ is the true complementary dynamics at step $i$.
Note that for this architecture it is equivalent to defining the loss based on MSE of the
\textit{total} dynamics. For weights $w_i$ we use a step profile:

\begin{equation}
	w_i = 
	\begin{cases}
		w_0 & 0 < i \leq p_t \\
		1 & p_t < i \leq p,
	\end{cases}
	\label{eq:archI_weight}
\end{equation} where $w_0 \ll 1$ is used to weight the first $p_t$ steps when the LSTM unit is still
under the transient effects of the cell states being initialized to zero. Predictions made during
this period is therefore valued much less. In practice, $p_t$ is usually negatively correlated
with the parametrization power of the reduction subspace $Y$ and can be determined empirically. For
optimization we use the gradient-based Adam optimizer \cite{KingmaB14} (also described in \nameref{S1_appendix}) with early stopping. The gradient is calculated for small batches of data series (batch
size $n_\text{batch}$) and across the entire training data for $n_\text{ep}$ epochs.

A notable property of this model architecture is that input representing the reduced state is always
accurate regardless of any errors made in predicting the dynamics previously. This is undesirable
especially for chaotic systems where errors tend to grow exponentially. Ideally, the model should be
optimized with respect to the cumulative effects of the prediction errors. To this end, this
architecture is primarily used for pre-training and a second architecture is utilized for fine-tuning
and multi-step-ahead prediction.

\subsubsection*{Architecture II}

The second architecture bears resemblance to the sequence-to-sequence (seq2seq) models which have
been widely employed for natural language processing tasks \cite{seq2seq, britz17}. It consists of
two stages (illustrated in Fig~\ref{fig2} II): a set-up stage and a prediction stage.
The set-up stage has the same structure as architecture I, taking as input a uniformly spaced
sequence of $s$ reduced-space states which we call $\{\boldsymbol{\xi}_{-s+1},
\boldsymbol{\xi}_{-s+2} ..., \boldsymbol{\xi}_{0}\}$. No output, however, is produced until the
very last step. This stage acts as a spin-up such that zero initializations to the
LSTM memory no longer affects prediction of dynamics at the beginning of the next stage. The output of the set-up stage is a single prediction of the
complementary dynamics $\hat{\boldsymbol{\psi}}_{0}$ corresponding to the last state of the input
sequence and the ending LSTM memory states. This dynamics is combined with
$\mathbf{F}_{\boldsymbol{\xi}}(\boldsymbol{\xi}_{0})$ to give the total dynamics at
$\boldsymbol{\xi}_{0}$. The final state and dynamics are passed to an integrator to obtain the
first input state of the prediction stage $\hat{\boldsymbol{\xi}}_{1}$. During the prediction
stage, complementary dynamics is predicted iteratively based on the newest state prediction and the
LSTM memory content before combined with $\mathbf{F}_{\boldsymbol{\xi}}$ dynamics to generate the
total dynamics and subsequently the next state. After $p$ prediction steps, the output of the model
is obtained as a sequence of predicted states $\{\hat{\boldsymbol{\xi}}_{1},
\ldots,\hat{\boldsymbol{\xi}}_{p}\}$ and a sequence of complementary dynamics
$\{\hat{\boldsymbol{\psi}}_{1}, \ldots,\hat{\boldsymbol{\psi}}_{p}\}$.

For this architecture we define the loss function as
\begin{equation}
	L = \sum\limits_{i=1}^{p} w_i||\hat{\boldsymbol{\psi}}_i +
	\mathbf{F}_{\boldsymbol{\xi}}(\hat{\boldsymbol{\xi}}_i) - \dot{\boldsymbol{\xi}}_i||^2.
	\label{eq:archII_loss}
\end{equation} This definition is based on MSE of the \textit{total} dynamics so that the
model learns to `cooperate' with the projected dynamics $\mathbf{F}_{\boldsymbol{\xi}}$. For
weights we use an exponential profile:
\begin{equation}
	w_i = \gamma^{i-1}, \; 0 < i \leq p
	\label{eq:archII_weight}
\end{equation} where $0 < \gamma < 1$ is a pre-defined ratio of decay. This profile is designed to
counteract the exponentially growing nature of the errors in a chaotic system and prevent exploding
gradients. Similar to architecture I, training is performed in batches using the Adam algorithm.

Architecture II, in contrast with the architecture I, finishes reading the entire input sequence
before producing the prediction sequence. For this reason it is suitable for running
multi-step-ahead predictions. Both architectures, however, share the same set of trainable weights
used to estimate the complementary dynamics. Hence, we can utilize architecture I as a pre-training
facility for architecture II because it tends to have smaller gradients (as errors do not accumulate
over time steps) and thus faster convergence. {This idea is very similar to teacher forcing method
used to accelerate training (see \cite{teacher_force})}. On the other hand, architecture II is much
more sensitive to the weights. Gradients tend to be large and only small learning rates can be
afforded. For more efficient training, it is therefore beneficial to use architecture I to find a
set of weights that already work with reasonable precision and perform fine-tuning with
architecture II. In addition, the $p_t$ parameter for architecture I also provides a baseline for
the set-up stage length $s$ to be used for architecture II.

Another feature of architecture II is that the length of its prediction stage can be arbitrary.
Shorter length limits the extent to which errors can grow and renders the model easier to train.
In practice we make sequential improvements to the model weights by progressively
increasing the length $p$ of the prediction stage.

For convenience, the hyperparameters involved in each architecture are summarized in 
Table~\ref{tab:hyps}.

\begin{table}[!ht]
\centering
\caption{
{\bf Summary of hyperparameters for data-driven model architectures.}}
\begin{tabular}{c|c|c|c}
\thickhline
\multicolumn{1}{l|}{\bf Category}      & {\bf Symbol}    			& {\bf Hyperparameter} & {\bf Architecture}  \\
\thickhline
\multirow{2}{*}{Layers}   & $n_{\text{LSTM}}$     		
						  & number of hidden units, LSTM layer               
						   & I \& II       \\
                          & $n_{\text{FC}}$       		
                           & number of hidden units, fully connected layer               
                            & I \& II \\ \hline 
\multirow{3}{*}{Series}   & $s$         					
						  & number of time steps, set-up stage              
						   & II       \\
                          & $p$         					
                           & number of time steps, prediction stage              
                            & I \& II       \\ 
                          & $\tau$         				
                           & time step              
                            & I \& II	\\ \hline
\multirow{3}{*}{Loss} 	 & $p_t$    						
						  & length of transient (low-weight) period                
						   & I       \\
                          & $w_0$    					
                           & transient weight               
                            & I       \\
                          & $\gamma$ 					
                           & weight decay               
                            & II \\ \hline 
\multirow{3}{*}{Training} & $n_{\text{batch}}$    	
                            & batch size               
                             & I \& II       \\
                          & $n_{\text{ep}}$    			
                           & number of epochs               
                            & I \& II       \\
                          & $\eta, \beta_1, \beta_2$ 	
                           & learning rate and momentum control               
                            & I \& II       \\ \thickhline
\end{tabular}
\label{tab:hyps}
\end{table}

\subsection*{Fully data-driven modeling} 

Both of the proposed architectures can be easily adapted for a fully data-driven modeling
approach (see \cite{Vlachas18}): for architecture I the sequence of total dynamics
$\{\dot{\boldsymbol{\xi}}_1, \ldots, \dot{\boldsymbol{\xi}}_p\}$ is used as the training target in
place of the complementary dynamics and for architecture II the FC layer output is directly
integrated to generate the next state. Doing so changes the distribution of model targets and
implicitly forces the model to learn more. For comparison, we examine the performance of this fully
data-driven approach through the example applications in the following section.

\section*{Results and discussion}

\subsection*{A chaotic intermittent low-order atmospheric model}
\label{sec:CDV}

\begin{figure}[!h]
\centering
\includegraphics[width=\textwidth]{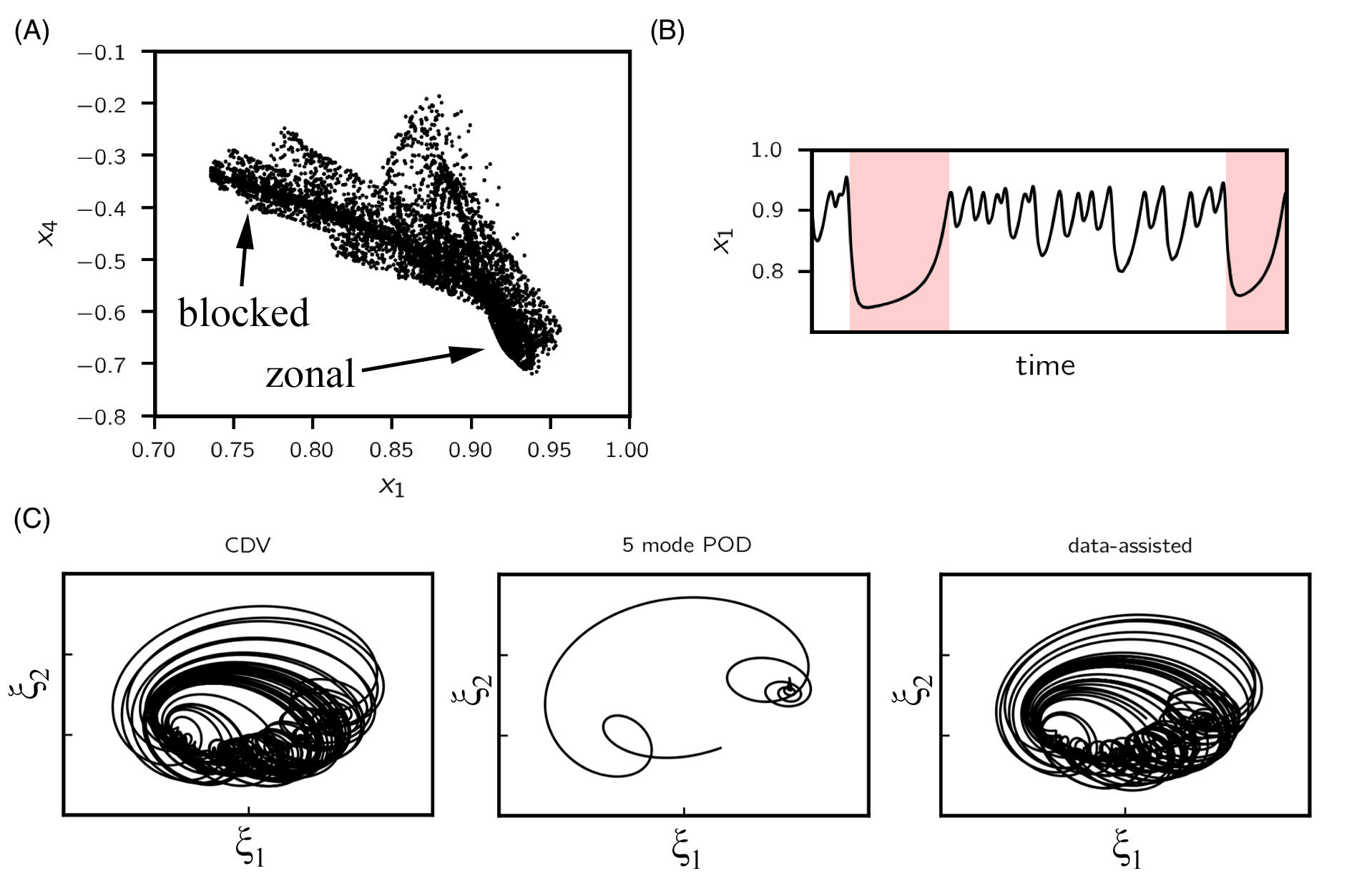}
\caption{{\bf CDV system.} (A) $10^4$ points sampled from the CDV attractor, projected to $(x_1, x_4)$ plane. (B) Example time series for $x_1$; \textit{blocked} flow regime is shaded in red. (C) Length-2000 trajectory projected to the first two POD modes (normalized) integrated using the CDV model (left), 5-mode POD projected model (middle) and data-assisted model (right). Despite preserving 99.6\% of the total variance, the 5-mode projected model has a single fixed point as opposed to a chaotic attractor. Data-assisted model, however, is able to preserve the geometric features of the original attractor.}
\label{fig3}
\end{figure}

We consider a chaotic intermittent low-order atmospheric model, the truncated Charney-DeVore (CDV) equations,
developed to model barotropic flow in a $\beta$-plane channel with orography. The model
formulation used herein is attributed  to \cite{crommelin04, Crommelin04b}, and employs a slightly different
scaling and a more general zonal forcing profile than the original CDV. Systems dynamics are
governed by the following ordinary differential equations:

\begin{equation} \label{eq:CDV} \begin{aligned}
	\dot{x}_1 &= \gamma_1^*x_3-C(x_1-x^*_1), &
	\dot{x}_2 &= -(\alpha_1x_1-\beta_1)x_3-Cx_2-\delta_1x_4x_6, \\
	\dot{x}_3 &= (\alpha_1x_1-\beta_1)x_2-\gamma_1x_1-Cx_3+\delta_1x_4x_5, &
	\dot{x}_4 &= \gamma^*_2x_6-C(x_4-x_4^*)+\varepsilon(x_2x_6-x_3x_5), \\
	\dot{x}_5 &= -(\alpha_2x_1-\beta_2)x_6-Cx_5-\delta_2x_4x_3, &
	\dot{x}_6 &= (\alpha_2x_1-\beta_2)x_5-\gamma_2x_4-Cx_6+\delta_2x_4x_2, \\
\end{aligned} \end{equation} where the model coefficients are given by

\begin{equation} \begin{aligned}
	\alpha_m &= \frac{8\sqrt{2}m^2(b^2+m^2-1)}{\pi(4m^2-1)(b^2+m^2)}, &
	\beta_m &= \frac{\beta b^2}{b^2+m^2}, \\
	\delta_m &= \frac{64\sqrt{2}}{15\pi}\frac{b^2-m^2+1}{b^2+m^2}, &
	\gamma_m^* &= \gamma\frac{4\sqrt{2}mb}{\pi(4m^2-1)}, \\
	\varepsilon &= \frac{16\sqrt{2}}{5\pi}, &
	\gamma_m &= \gamma\frac{4\sqrt{2}m^3b}{\pi(4m^2-1)(b^2+m^2)},
\end{aligned} \end{equation} for $m = 1,2$. Here we examine the system at a fixed set of parameters
$(x_1^*,x_4^*,C,\beta,\gamma,b) = (0.95,-0.76095,0.1,1.25,0.2,0.5)$, which is found to
demonstrate chaotic intermittent transitions between \textit{zonal} and \textit{blocked} flow regime, caused by
the combination of topographic and barotropic instabilities \cite{crommelin04, Crommelin04b}. These highly transient instabilities render this model an appropriate test case for evaluating the developed methodology. The two distinct regimes are manifested through $x_1$ and $x_4$ (Fig~\ref{fig3}A and \ref{fig3}B). 


For reduction of the system we attempt the classic proper orthogonal decomposition (POD) whose
details are described in \nameref{S1_appendix}. The basis vectors of the projection subspace are
calculated using the method of snapshots on a uniformly sampled time series of length 10,000 obtained
by integrating Eq~(\ref{eq:CDV}).  The first five POD modes
collectively account for 99.6\% of the total energy. However, despite providing respectable
short-term prediction accuracy, projecting the CDV system to its most energetic five modes completely changes
the dynamical behavior and results in a single globally attracting fixed point instead of a strange
attractor. The difference between exact and projected dynamics can be seen in terms of the two most energetic POD coefficients, $\xi_1, \xi_2$, in Fig~\ref{fig3}C (left and middle subplots).


In the context of our framework, we construct a data-assisted reduced-order model that includes the
dynamics given by the 5-mode POD projection. We set $n_{\text{LSTM}}=1$ (because one dimension is
truncated) and $n_{\text{FC}}=16$. Input to the FC layer is a concatenation of LSTM output and
reduced state because $n_{\text{LSTM}}=1$ is sufficient to represent the truncated mode. Data is
obtained  as 10,000 trajectories, each with $p = 200$ and $\tau = 0.01$. We use 80\%, 10\%, 10\%
for training, validation and testing respectively. For this setup it proves sufficient, based on
empirical evidence, to train the assisting data-driven model with Architecture I for 1000 epochs,
using a batch size of 250. The trained weights are plugged in architecture II to generate
sequential predictions. As we quantify next, it is observed that (a) the trajectories behave much
like the 6-dimensional CDV system in the long term by forming a similar attractor, as shown in
Fig~\ref{fig3}C, and (b) the short-term prediction skill is boosted significantly.

\begin{figure}[!b]
\centering
\includegraphics[width=\textwidth]{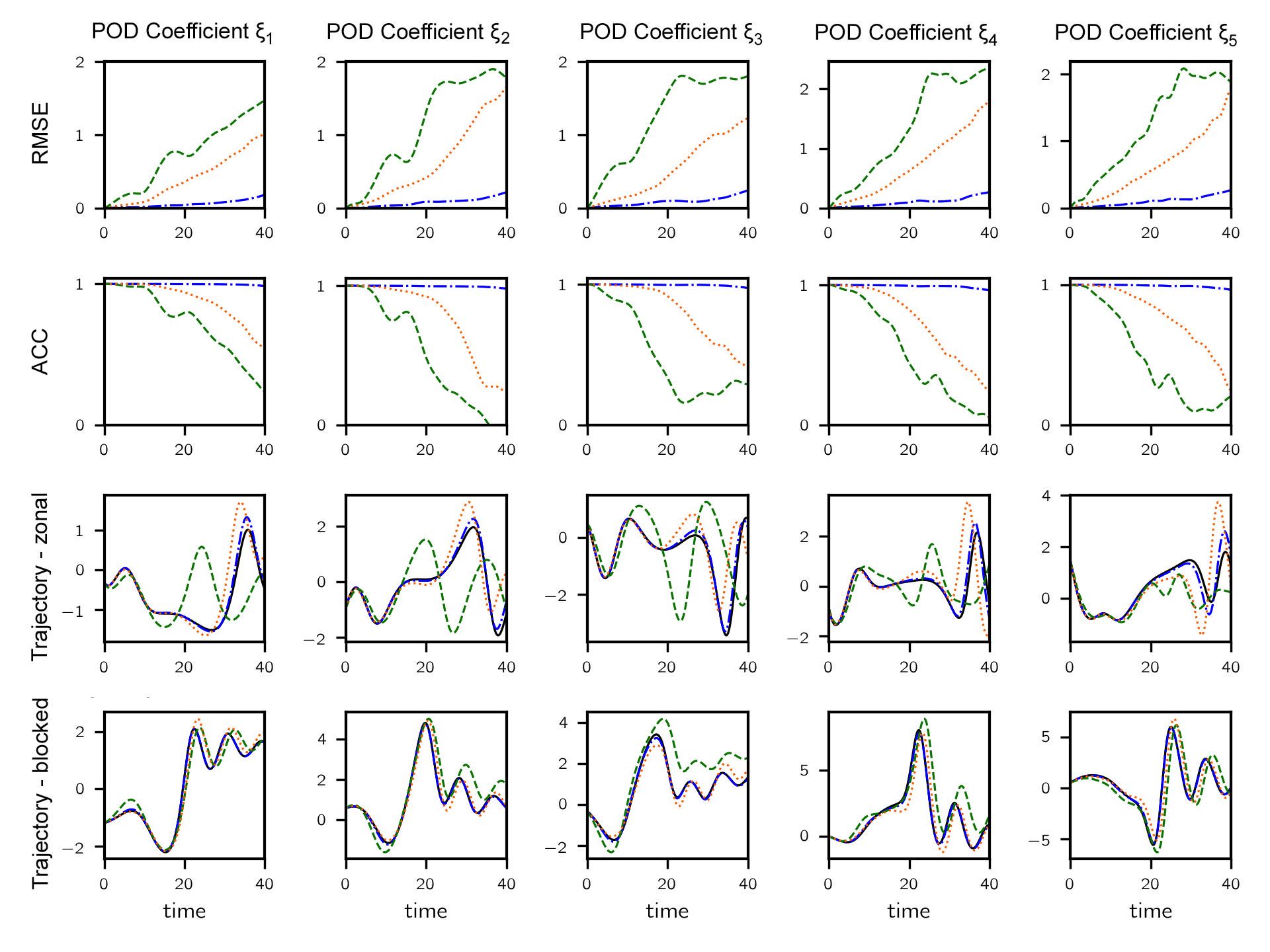}
\caption{{\bf Results for CDV system.} (Row 1) RMSE vs. lead time for 5-mode POD projected model (\protect\leqn), data-assisted model (\protect\lhybrid) and purely data-driven model (\protect\ldata). (Row 2) ACC vs. lead time. (Row 3) A sample trajectory corresponding to zonal flow - true trajectory is shown (\protect\ltruth). (Row 4) A sample trajectory involving regime transition (happening around $t = 20$) - true trajectory is shown (\protect\ltruth). For rows 1, 3 and 4, plotted values are normalized by the standard deviation of each dimension.}
\label{fig4}
\end{figure}

We  quantify the improvement in prediction performance by using two error metrics - root
mean squared error (RMSE) and correlation coefficient. For comparison we also include prediction
errors when using a purely data-driven model based on LSTM. RMSE in $i$th reduced dimension is
computed as

\begin{equation}
	\text{RMSE}_i(t_l) = \sqrt{\frac{1}{N}\sum\limits_{n=1}^N \left(\xi^{(n)}_{i}(t_l) -
		\hat{\xi}^{(n)}_{i}(t_l)\right)^2}, \;\; i = 1,...,m,
	\label{eq:rmse}
\end{equation} where $\xi_i^{(n)}(t_l)$ and $\hat{\xi}_i^{(n)}(t_l)$ represent the truth and
prediction for the $n$th \textit{test} trajectory at prediction lead time $t_l$ respectively. The
results are plotted in Fig~\ref{fig4}. We end remark that the predictions obtained by
the proposed data-assisted model are significantly better than the projected model, as well as than
the purely data-driven approach. Low error levels are maintained by the present approach even when
the other methods under consideration exhibit  significant errors.


The anomaly correlation coefficient (ACC) \cite{Jolliffe11} measures the correlation between
anomalies of forecasts and those of the truth with respect to a reference level and is defined as
\begin{equation}
	\text{ACC}_i(t_l) = \frac{\sum\limits_{n=1}^N \left(\xi^{(n)}_{i}(t_l) -
		\overline{\xi}_i\right)\left(\hat{\xi}_i^{(n)}(t_l) - \overline{\xi}_i\right)}
		{\sqrt{\sum\limits_{n=1}^N \left(\xi^{(n)}_{i}(t_l) - \overline{\xi}_i\right)^2
		\sum\limits_{n=1}^N \left(\hat{\xi}^{(n)}_{i}(t_l) - \overline{\xi}_i\right)^2}} 
\end{equation} where $\overline{\xi}_i$ is the reference level set to the observation
average by default. ACC takes a maximum value of 1 if the variation pattern of the anomalies of
forecast is perfectly coincident with that of truth and a minimum value of -1 if the pattern is
completely reversed. Again, the proposed method is able to predict anomaly variation patterns which
are almost perfectly correlated with the truth at very large lead times when the predictions made
by the compared methods are mostly uncorrelated (Fig~\ref{fig4} - second row).

In the third and fourth rows of Fig~\ref{fig4} we illustrate the improvement that we
obtain with the data-assisted approach throughout the systems attractor, i.e. in both zonal and
blocked regimes. In the third row of Fig~\ref{fig4} the flow in the zonal regime is
shown and in the fourth row we demonstrate the flow transitions into the blocked regime around
$t=20$. In both cases, the data-assisted version clearly  improves the prediction accuracy.

We emphasize that the presence of the equation-driven part contributes largely to the
long-term stability (vs. purely data driven models) while the data-driven part serves to
improve the short-term prediction accuracy. These two ingredients of the dynamics complement each other
favorably in achieving great prediction performance. In addition, the data-assisted approach
successfully produces a chaotic structure that is similar to the one observed in
the full-dimensional system, a feat that cannot be replicated by either methods using equation or data
alone.


\subsection*{Intermittent bursts of dissipation in Kolmogorov flow}
\label{sec:kol2D}


We consider the two-dimensional incompressible Navier-Stokes equations
\begin{equation} 
	\begin{gathered}
		\partial_t\mathbf{u} = -\mathbf{u}\cdot\nabla\mathbf{u} - \nabla p +\nu\Delta\mathbf{u} +
			\mathbf{f} \\
		\nabla \cdot \mathbf{u} = 0
	\end{gathered} 
	\label{eq:kol2D} 
\end{equation} where $\mathbf{u} = (u_x, u_y)$ is the fluid velocity defined over the domain
$(x,y)\in\Omega = [0, 2\pi]\times[0, 2\pi]$ with periodic boundary conditions, $\nu = 1/Re$ is the
non-dimensional viscosity equal to reciprocal of the Reynolds number and $p$ denotes
the pressure field over $\Omega$. We consider the flow driven by the monochromatic Kolmogorov
forcing $\mathbf{f}(\mathbf{x}) = (f_x, f_y)$ with $f_x = \sin(k_fy)$ and $f_y = 0$. $\mathbf{k}_f
= (0, k_f)$ is the forcing wavenumber.

Following \cite{Faraz17}, the kinetic energy $E$, dissipation $D$ and energy input $I$ are defined
as
\begin{equation} \begin{gathered}
	E(\mathbf{u}) = \frac{1}{|\Omega|}\int_\Omega\frac{1}{2}|\mathbf{u}|^2\;d\Omega, \\
	D(\mathbf{u}) = \frac{\nu}{|\Omega|}\int_\Omega|\nabla u|^2\;d\Omega, \\
	I(\mathbf{u}) = \frac{1}{|\Omega|}\int_\Omega\mathbf{u}\cdot\mathbf{f}\;d\Omega
\end{gathered}\end{equation} satisfying the relationship $\dot{E} = I - D$. Here $|\Omega| =
(2\pi)^2$ denotes the area of the domain.

\begin{figure}[!h]
\centering
\includegraphics[width=\textwidth]{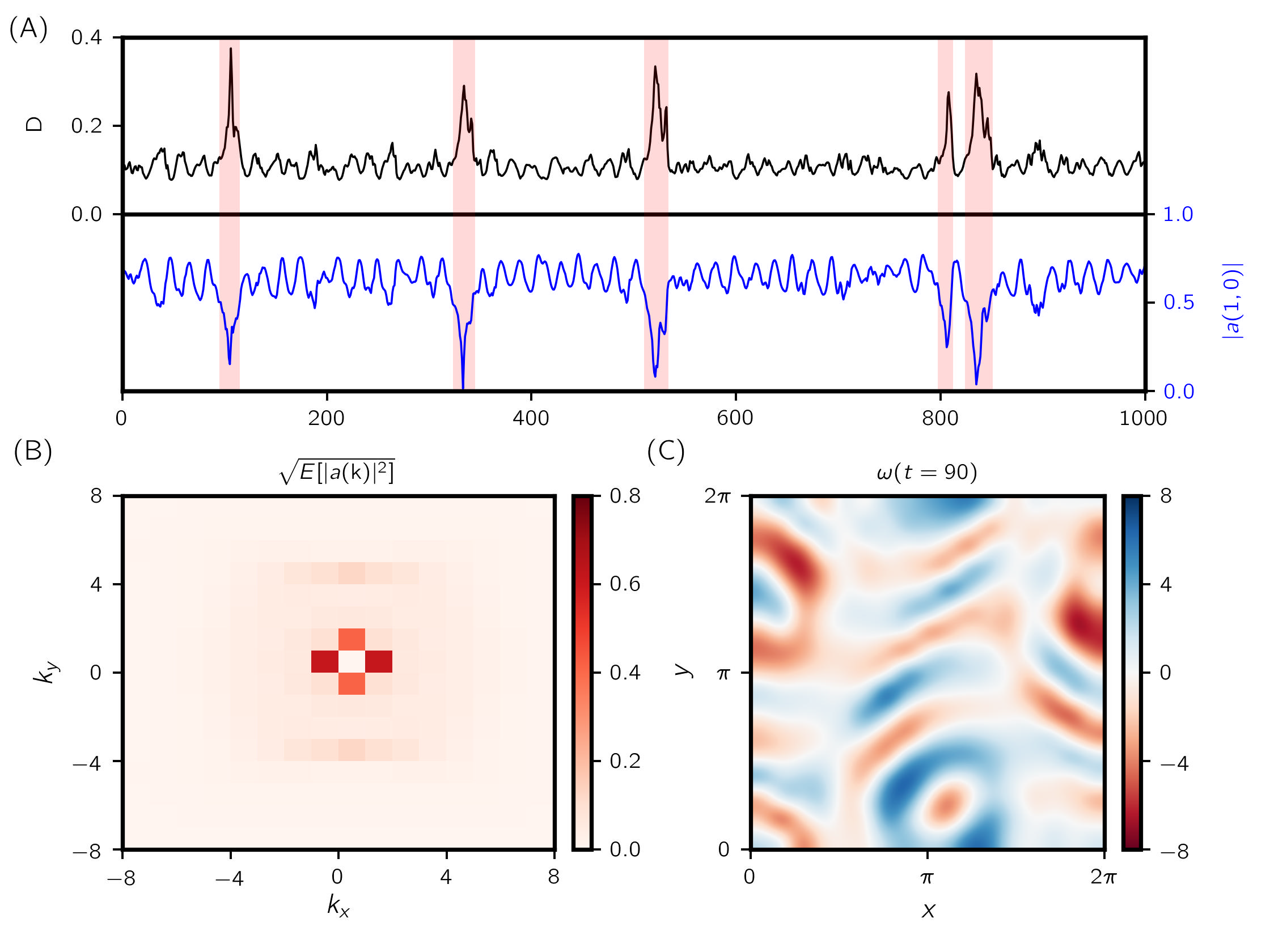}
\caption{{\bf Kolmogorov flow.} (A) Time series of energy dissipation rate $D$ and Fourier coefficient modulus $|a(1,0)|$ - rare events are signaled by burst in $D$ and sudden dip in $|a(1,0)|$. (B) Root mean squared (RMS) modulus for wavenumbers $-8 \leq k_1, k_2 \leq 8$.  (C) Vorticity field $\nabla\times\mathbf{u} = \omega$ at time $t = 90$ over the domain $\mathbf{x} \in [0,2\pi]\times[0, 2\pi]$}
\label{fig5}
\end{figure}

The Kolmogorov flow admits a laminar solution $u_x = (Re/k_f^2)\sin(k_fy), u_y = 0$. For
sufficiently large $k_f$ and $Re$, this laminar solution is unstable, chaotic and exhibiting
intermittent surges in energy input $I$ and dissipation $D$. Here we study the flow under a
particular set of parameters $Re = 40$ and $k_f = 4$ for which we have the occurrence of extreme events. Fig~ \ref{fig5}A shows the bursting time series of the dissipation $D$ along a sample trajectory.


Due to spatial periodicity, it is natural to examine the velocity field in Fourier space. The
divergence-free velocity field $\mathbf{u}$ admits the following Fourier series expansion:

\begin{equation}
	\mathbf{u}(\mathbf{x},t) = \sum\limits_{\mathbf{k}} \frac{a(\mathbf{k},t)}{|\mathbf{k}|}
	\begin{pmatrix}
		k_2 \\ -k_1
	\end{pmatrix} e^{i\mathbf{k}\cdot\mathbf{x}}
	\label{eq:kol2D_fourier}
\end{equation} where $\mathbf{k}=(k_1, k_2)$ is the wavenumber and $a(\mathbf{k},t) =
-\overline{a(-\mathbf{k},t)}$ for $\mathbf{u}$ to be real-valued. For notation clarity, we will not
explicitly write out the dependence on $t$ from here on. Substituting Eq~(\ref{eq:kol2D_fourier}) into
the governing equations Eq~(\ref{eq:kol2D}) we obtain the evolution equations for $a$ as (more details are presented in \nameref{S1_appendix})

\begin{equation}
	\dot{a}(\mathbf{k}) = \sum\limits_{\mathbf{p}+\mathbf{q}=\mathbf{k}} i\frac{(p_1q_2-p_2q_1)(k_1q_1+k_2q_2)}{|\mathbf{p}||\mathbf{q}||\mathbf{k}|}a(\mathbf{p})a(\mathbf{q}) -\nu|\mathbf{k}|^2a(\mathbf{k}) -\frac{1}{2}i(\delta_{\mathbf{k},\mathbf{k}_f} + \delta_{\mathbf{k},-\mathbf{k}_f}) 
	\label{eq:triad_dynamics}
\end{equation}

The first term suggests that any mode with wavenumber $\mathbf{k}$ is directly affected, in a
nonlinear fashion, by pairs of modes with wavenumbers $\mathbf{p}$ and $\mathbf{q}$ such that
$\mathbf{k} = \mathbf{p} + \mathbf{q}$. A triplet of modes $\{\mathbf{p}, \mathbf{q}, \mathbf{k}\}$
satisfying this condition is referred to as a \textit{triad}. It is worth noting that a
mode which does not form a triad with mode $\mathbf{k}$ can still have an indirect effect on
dynamics $\dot{a}(\mathbf{k})$ through interacting with modes that do form a triad with
$\mathbf{k}$.

In \cite{Faraz17}, it is found that the most revealing triad interaction to observe, in the interest
of predicting intermittent bursts in the energy input/dissipation, is amongst modes $(0, k_f), (1,
0)$ and $(1,k_f)$. Shortly prior to an intermittent event, mode $(1, 0)$ transfers a large amount of
energy to mode $(0, k_f)$, leading to rapid growth in the energy input rate $I$ and subsequently the
dissipation rate $D$ (see Fig~\ref{fig5}A). However, projecting the velocity field
and dynamics to this triad of modes and their complex conjugates fails to faithfully replicate the
dynamical behaviors of the full system (the triad only accounts for 59\% of the total energy;
Fourier energy spectrum is shown in Fig~\ref{fig5}B). We use the present 
framework to complement the projected triad dynamics.


Quantities included in the model are $a(1,0), a(0, k_f)$ and $a(1, k_f)$ and their conjugate pairs,
which amount to a total of six independent dimensions. Data is generated as a single time series of
length $10^5$ at $\Delta t=1$ intervals, by integrating the full model equations Eq~(\ref{eq:kol2D})
using a spectral grid of size 32 (wavenumbers truncated to $-16 \leq k_1, k_2 \leq 16$)
\cite{Uecker09}. Each data point is then taken as an initial condition from which a trajectory of
length 1 (200 steps of 0.005) is obtained. The sequence of states along the trajectory is projected
to make up the 6-dimensional input to the LSTM model. The ground truth total dynamics is again
approximated with first-order finite differences. The first 80\% of the data is used for training,
5\% for validation and the remaining 15\% for testing.

For this problem it is difficult to compute the true minimum parametrizing dimension so we
conservatively choose $n_{\text{LSTM}} = 70$ and $n_{\text{FC}} = 38$. It is found that the models
do not tend to overfit, nor is their performance sensitive to these hyper-parameters around the
chosen value. Since the number of hidden units used in the LSTM is large relative to the input
dimension, they are not concatenated with the input before entering the output layer. We first
perform pre-training with architecture I for 1000 epochs and fine-tune the weights with
architecture II. Due to the low-energy nature of the reduction space, transient effects are
prominent (see \nameref{S1_appendix}) and thus a sizable set-up stage is needed for training
and prediction with architecture II. Using a sequential training strategy, we keep $s=100$ fixed
and progressively increase prediction length at $p = \{10, 30, 50, 100\}$ (see \nameref{S1_appendix}). At each step, weights are optimized for 1000 epochs using a batch size of
250. The hyperparameters defining the loss functions are $p_t = 60$, $w_0 = 0.01$ and $\gamma =
0.98$, which are found empirically to result in favorable weight convergence.


Similar to the CDV system, we measure the prediction performance using RMSE and correlation
coefficient. Since the modeled Fourier coefficients are complex-valued, the sum in
Eq~(\ref{eq:rmse}) is performed on the squared complex magnitude of the absolute error.

\begin{figure}[!h]
\centering
\includegraphics[width=\textwidth]{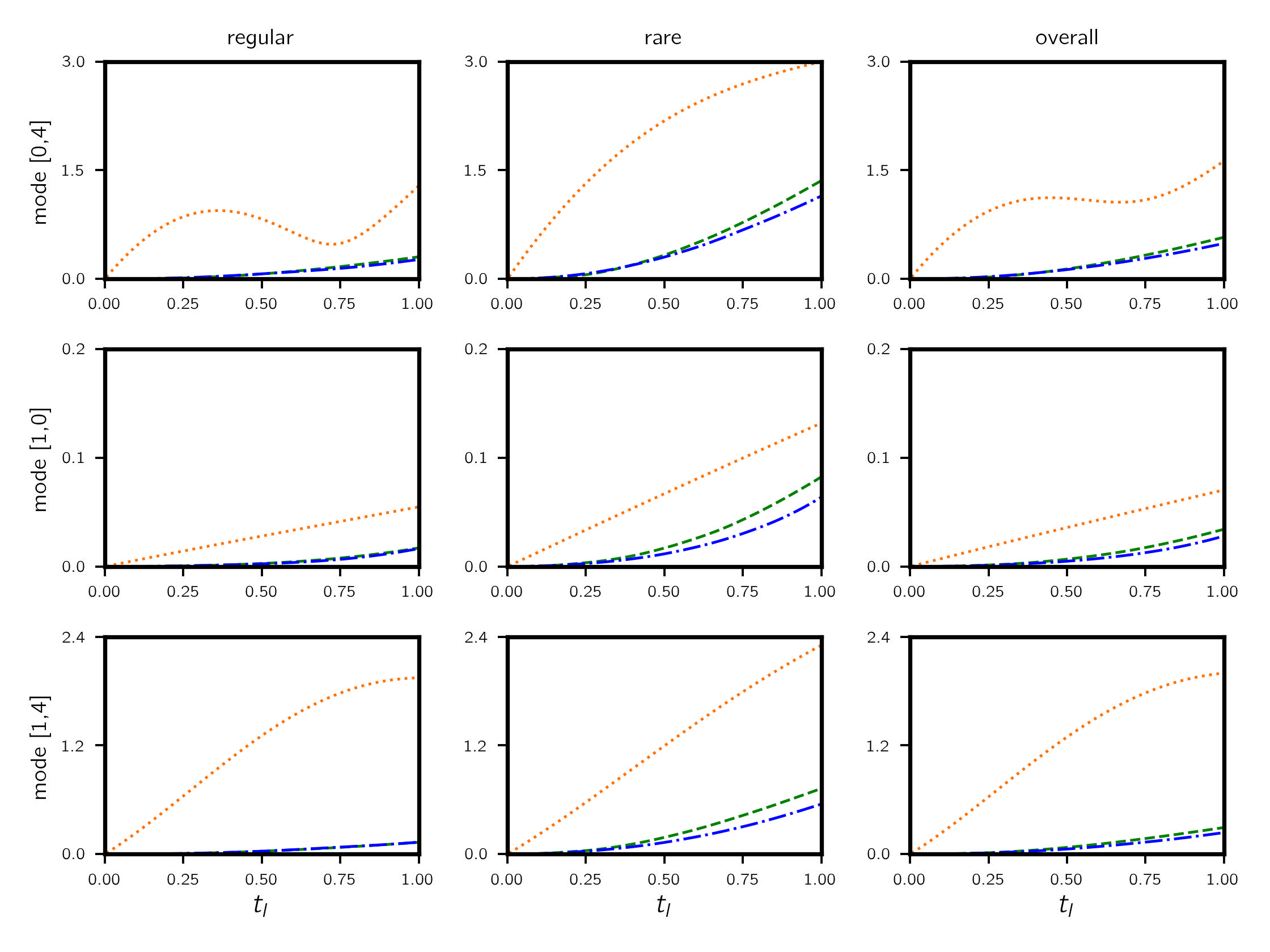}
\caption{{\bf Kolmogorov flow - RMSE vs. time.} Errors are computed for $10^4$ test trajectories (legend: fully data-driven \protect\ldata; data-assisted \protect\lhybrid; triad \protect\leqn). The RMSE in each mode is normalized by the corresponding amplitude $E(\mathbf{k}) = \sqrt{\mathbb{E}[|a(\mathbf{k})|^2]}$. A test trajectory is classified as \textit{regular} if $|a(1,0)|>0.4$ at $t=0$ and \textit{rare} otherwise. Performance for regular, rare and all trajectories are shown in three columns. Data-assisted model has very similar errors to those of purely data-driven models for regular trajectories, but the performance is visibly improved for rare events.}
\label{fig6}
\end{figure}

\begin{figure}[!h]
\centering
\includegraphics[width=\textwidth]{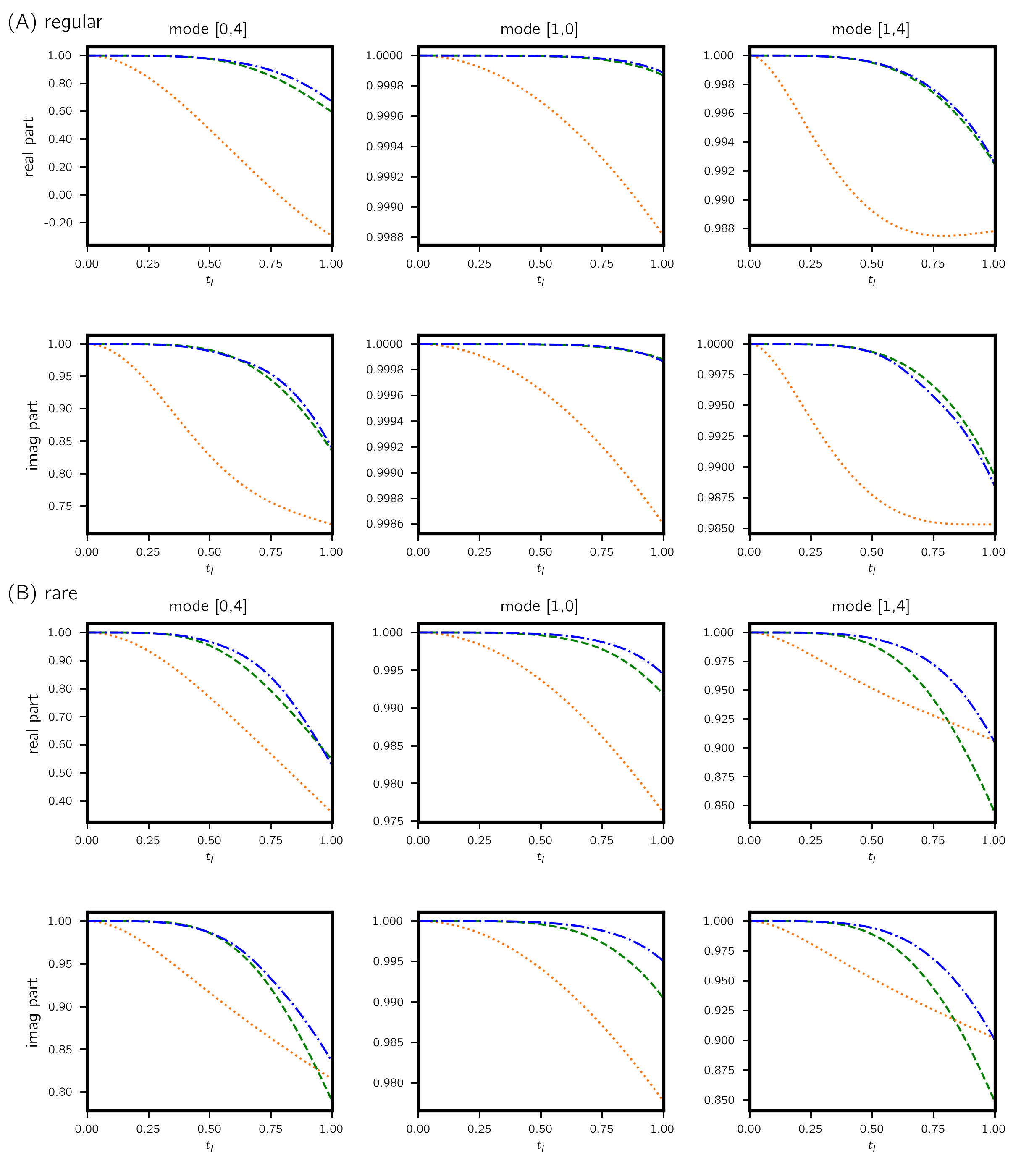}
\caption{{\bf Kolmogorov flow: ACC vs. time.} Values are computed for (A) regular and (B) rare trajectories classified from $10^4$ test cases. Legend: fully data-driven \protect\ldata; data-assisted \protect\lhybrid; triad dynamics \protect\leqn. Real and imaginary parts are treated independently. Similarly to RMSE in Fig~\ref{fig6}, improvements in predictions made by the data-assisted model are more prominent for rare events.}
\label{fig7}
\end{figure}

\begin{figure}[!h]
\centering
\includegraphics[width=\textwidth]{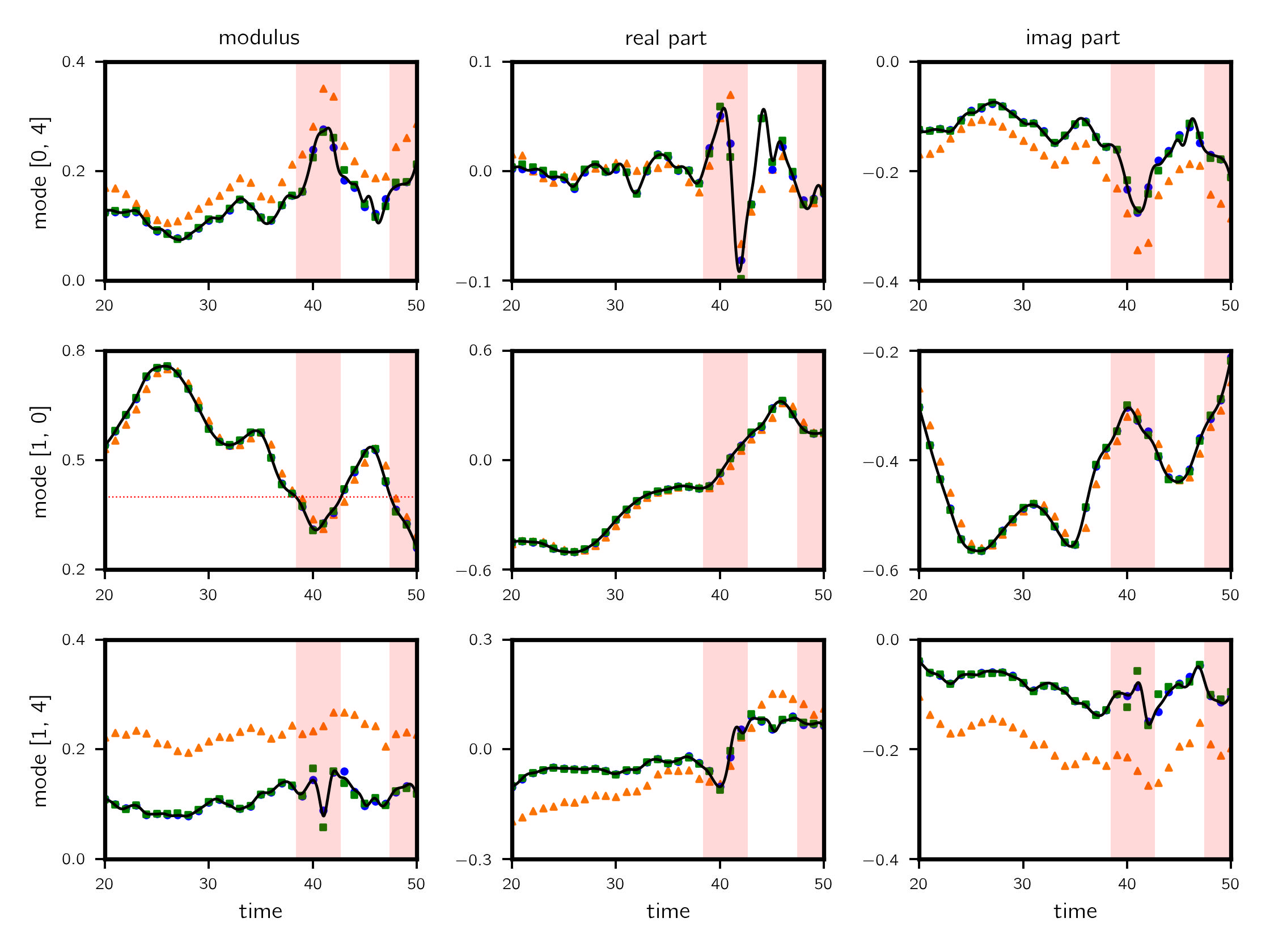}
\caption{{\bf Kolmogorov flow: predictions along a sample trajectory with lead time = 0.5.}  Results for the complex
	modulus (left column), real part (middle column) and imaginary part (right column) of the
	wavenumber triad are shown. Legend: truth \protect\ltruth; data-assisted \protect\phybrid; triad
	dynamics \protect\peqn; purely data-driven \protect\pdata. Rare events are recorded when $|a(1,0)|$
	(left column, mid row) falls below 0.4 (shaded in red). Significant improvements are observed for
	wavenumbers $(0,4)$ and $(1,4)$.}
\label{fig8}
\end{figure}

The resulting normalized test error curves are shown in Figs~\ref{fig6} and
\ref{fig7} respectively, comparing the proposed data-assisted framework with the original
projected model and the fully data-driven approach as the prediction lead time increases. At 0.5
lead time (approximately 1 eddy turn-over time $t_e$), the data-assisted approach achieves 0.13,
0.005, 0.058 RMSE in mode $[0,4]$, $[1,0]$ and $[1,4]$ respectively. Predictions along a sample
trajectory is shown in Fig~\ref{fig8}.

Overall, the data-assisted approach produces the lowest error, albeit narrowly beating the fully
data-driven model but significantly outperforming the projected model (88\%, 86\% and 95\%
reduction in error for the three modes). This is because data is used to assist a projected model
that ignores a considerable amount of state information which contribute heavily to the dynamics.
It is therefore all up to the data-driven model to learn this missing information. For this reason
we observe similar performance between data-assisted and fully data-driven models. However, when we
classify the test cases into regular and rare events based on the value of $|a(1,0)|$ and examine
the error performance separately, the advantage of the data-assisted approach is evident in the
latter category, especially for mode $(1,4)$. The is mainly due to (a) rare events appear less
frequently in data such that the corresponding dynamics is not learned well compared to regular
events and (b) the triad of Fourier modes selected play more prominent role in rare events and therefore
the projected dynamics contain relevant dynamical information. Nevertheless, errors for rare events are
visibly higher (about 5 times), attesting to their unpredictable nature in general.

\begin{figure}[!h]
\centering
\includegraphics[width=\linewidth]{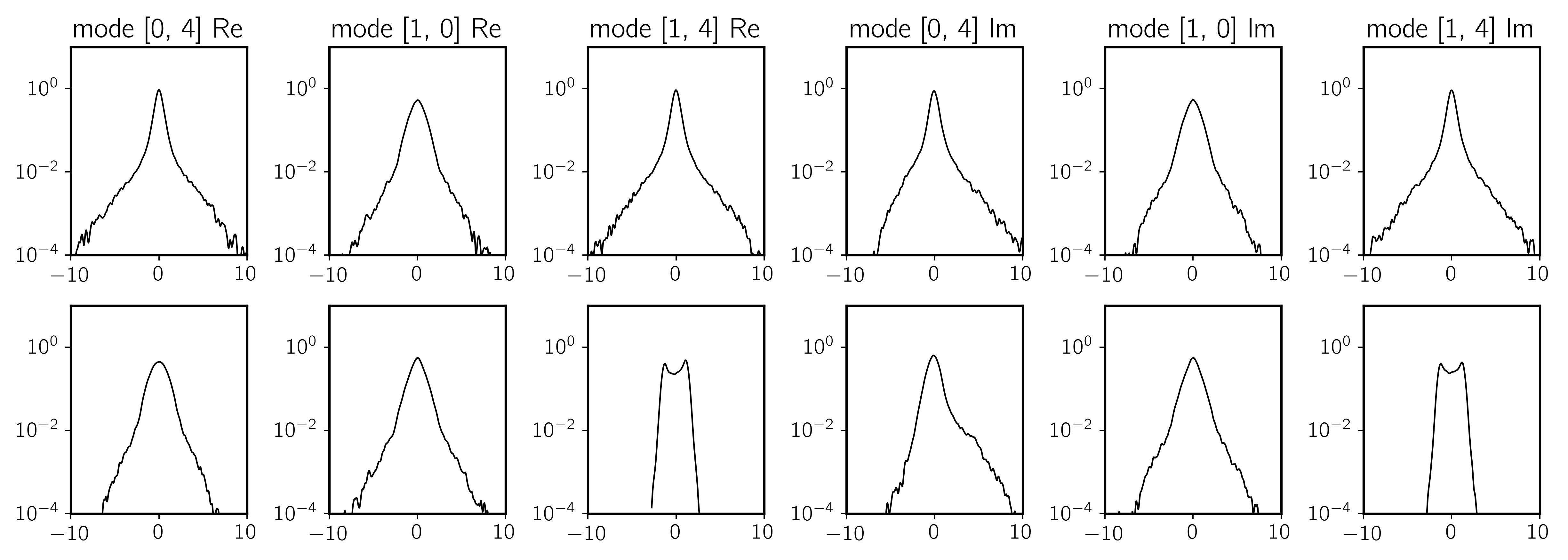}
\caption{{\bf Marginal probability density function of total dynamics (top row) and complementary dynamics (bottom row).} Horizontal axes are scaled by standard deviations of each quantity. For the real and imaginary parts of mode $(1, 4)$ (and real part of mode $(0, 4)$ to a smaller degree) triad dynamics help remove large-deviation observations in the complementary dynamics.}
\label{fig9}
\end{figure}

To better understand the favorable properties of the hybrid  scheme when it comes to the prediction
of extreme events we plot the probability density function (pdf) of complementary and total
dynamics (see eq. Eq~(\ref{eq:exact_red_dynamics})), calculated from the $10^5$-point training data set with a kernel density estimator
(Fig~\ref{fig9}). The dynamics values are standardized so that 1 unit in horizontal
axis represent 1 standard deviation. We immediately notice that total dynamics in every dimension
have a fat-tailed distribution. This signifies  that the data set contains several \textit{extreme}
observations, more than 10 standard deviations away from the mean. For data-driven models these
dynamics are difficult to learn due to their sporadic occurrence in sample data and low density in
phase space.


In contrast, the marginal pdf of the complementary dynamics have noticeably different
characteristics, especially in both the real and imaginary parts of mode $(1, 4)$ (and to a smaller
degree for the real part of mode $(0, 4)$). The distribution is bimodal-like; more importantly, density
falls below $10^{-4}$ level within 3 standard deviations. Because of its concentrated character, this is a much better conditioned target
distribution, as the data-driven scheme would never have to learn extreme event dynamics; these are captured by the projected equations. As expected, the error plot in Fig~\ref{fig7} suggests that
the biggest improvement from a purely data-driven approach to a data-assisted approach is indeed
for mode $(1, 4)$.

\section*{Conclusion}
We introduce a data assisted  framework for reduced-order modeling and prediction of extreme transient events in complex dynamical systems with high-dimensional attractors. The framework utilizes a data-driven approach to complement the dynamics given by
imperfect models obtained through projection, i.e. in cases when the projection subspace does not
perfectly parametrize the inertial manifold of the system. Information which is invisible to the
subspace but important to the dynamics is extracted by analyzing the time history of trajectories (data-streams) projected in the subspace, using a RNN strategy. The LSTM based architecture of the employed RNN allows for the modeling of the dynamics using delayed coordinates, a feature that significantly improves the performance of the scheme, complementing observation in fully data-driven schemes.

We showcase the capabilities of the present approach through two illustrative examples exhibiting intermittent bursts: a low dimensional atmospheric system, the Charney-DeVore model and a high-dimensional system, the Kolmogorov flow described by Navier-Stokes equations. For the former the data-driven model helps to improve significantly  the short-term prediction skill in a high-energy reduction subspace, while faithfully replicating the chaotic attractor of the original system. In the infinite-dimensional example it is clearly demonstrated that in regions characterized by extreme events the data-assisted strategy is more effective than the fully data-driven prediction or the projected equations. On the other hand, when we consider the performance close to the main attractor of the dynamical system the purely data-driven approach and the data-assisted scheme exhibit comparable accuracy. 

The present approach provides a non-parametric framework for the improvement of imperfect models through data-streams. For regions where data is available we obtain corrections for the model, while for regions where no data is available the underlying model still provides a baseline for prediction. The results in this work emphasize the value of this hybrid  strategy for the prediction of extreme transient responses for which data-streams may not contain enough information. In the examples considered the imperfect models where obtained through projection to low-dimensional subspaces. It is important to emphasize that such imperfect models should contain relevant dynamical information for the modes associated to extreme events. These modes are not always the most energetic modes (as illustrated in the fluids example) and numerous efforts have been devoted for their characterization \cite{Faraz17, cousinsSapsis2015_JFM, Farazmand2016}. 

Apart of the modeling of extreme events, the developed blended strategy should be of interest for data-driven modeling of systems exhibiting singularities or singular perturbation problems. In this case the governing equations have one component that is particularly challenging to model with data, due to its singular nature. For such systems it is beneficial to combine the singular part of the equation with a data-driven scheme that will incorporate information from data-streams. Future work will focus on the application of the formulated method in the context of predictive
control \cite{slotine92, brunton16a, mezic18} for turbulent fluid flows and in particular for the
suppression of extreme events.

\section*{Supporting information}


\paragraph*{S1 Appendix.}
\label{S1_appendix}
{\bf Supplementary notes.} In the notes we provide some background theory on recurrent neural networks, LSTM and momentum based optimization methods. Additional computation results for CDV and Kolmogorov flow are also included.

\paragraph*{S2 Code.}
\label{S2_code}
{\bf Python source code.} All code used in this study is available at: \href{https://github.com/zhong1wan/data-assisted}{https://github.com/zhong1wan/data-assisted}. In addition, all training and testing data files are available from the authors upon request.

\section*{Acknowledgments}
TPS and ZYW have been supported through the AFOSR-YIA FA9550-16-1-0231, the ARO MURI W911NF-17-1-0306 and the ONR MURI N00014-17-1-2676. PK and PV acknowledge support by the Advanced Investigator Award of the European Research Council (Grant No: 341117). 


\bibliography{bibs}

\end{document}